\documentclass[prd,nofootinbib,english]{revtex4}
\usepackage{amsmath}
\usepackage{amsfonts,color}
\usepackage{amsmath}
\usepackage{amsthm}
\usepackage{pdfpages}
\usepackage{amsfonts,color}
\usepackage{amssymb,float}

\usepackage[utf8]{inputenc}
\allowdisplaybreaks

\usepackage{color}
\usepackage{hyperref}
\hypersetup{
    colorlinks=true,
    linkcolor=blue,
    filecolor=magenta,      
   citecolor=blue
}

\usepackage{enumitem}

\usepackage{tikz}
\usetikzlibrary{shapes.geometric}
\usetikzlibrary{arrows.meta,arrows}

\begin{document}
\title{Observational Constraints in Metric-Affine Gravity}

\author{Sebastian Bahamonde$^{1,2,}$}
\email{sbahamonde@ut.ee, sebastian.beltran.14@ucl.ac.uk}
\author{Jorge Gigante Valcarcel$^{1,}$}
\email{jorge.gigante.valcarcel@ut.ee}
\affiliation{$^{1}$Laboratory of Theoretical Physics, Institute of Physics, University of Tartu, W. Ostwaldi 1, 50411 Tartu, Estonia \\
$^{2}$Laboratory for Theoretical Cosmology, Tomsk State University of
Control Systems and Radioelectronics, 634050 Tomsk, Russia (TUSUR)}

\begin{abstract}

We derive the main classical gravitational tests for a recently found vacuum solution with spin and dilation charges in the framework of Metric-Affine gauge theory of gravity. Using the results of the perihelion precession of the star S2 by the GRAVITY collaboration and the gravitational redshift of Sirius B white dwarf we constrain the corrections provided by the torsion and nonmetricity fields for these effects.

\end{abstract}

\maketitle

\section{Introduction}

Since its inceptions, General Relativity (GR) has been established as the most accurate and successfully tested theory of gravity. Some of its most important experimental predictions include the perihelion shift of astrophysical bodies, the deflection of light, the gravitational redshift, the time dilation and the detection of gravitational waves, among others \cite{Will:2014kxa,Abbott:2016blz,TheLIGOScientific:2017qsa}. Nevertheless, there are still fundamental issues to match the observations especially at galactic and cosmological scales, where the requirement of dark matter and energy components is still an open-ended aspect to clarify. In any case, further constraints involving modifications of the gravitational field can also be obtained at lower energies, as in the vicinity of the solar system or close to other astrophysical compact objects where the standard framework of GR is significantly well tested.

In the framework of extended theories of gravity, the introduction of post-Riemannian degrees of freedom into the geometrical structure of the space-time displays new properties of the gravitational field. In particular, the presence of nonvanishing torsion and nonmetricity components in the affine connection can be related to the intrinsic hypermomentum of matter and provide significant corrections to the standard approach of GR, even in the Riemannian sector governed by the metric structure. From a theoretical point of view, the resulting geometry can be related to the existence of a new fundamental symmetry by applying the gauge principles, which leads to the appearance of the Metric-Affine Gauge theory of gravity (MAG) \cite{Hehl:1994ue,Blagojevic:2013xpa,Cabral:2020fax}.

The existing correspondence between these geometrical quantities and their matter sources with microstructure clearly points out that the underlying corrections differ for the case of ordinary matter uncoupled to the general affine connection and for those particles with intrinsic hypermomentum \cite{Obukhov:1993pt,Obukhov:1996mg,Iosifidis:2020gth}. In any case, the dynamical character of the connection enables the existence of fully relativistic configurations with independent torsion and nonmetricity fields, which can be induced on the metric tensor and thereby operate on the geodesic motion of ordinary matter via the Levi-Civita connection~\cite{Bahamonde:2020fnq,Cembranos:2016gdt,Cembranos:2017pcs}. In this work, we consider recent observations of the perihelion shift of the star S2 around Sagittarius A* (Sgr A*) and the gravitational redshift of Sirius B to constrain the dynamical effects of torsion and nonmetricity under this approach.

For this task, we organise this paper as follows. In Sec. \ref{sec:metricaffine}, we briefly present the foundations of metric-affine geometry and revisit the recent exact solution with independent dynamical torsion and nonmetricity fields found in \cite{Bahamonde:2020fnq}. Indeed, this solution constitutes an isolated gravitational system characterised by a metric tensor with spin and dilation charges and accordingly allows us to study the phenomenological contributions of both charges in a unified way under the aforementioned approach. We compute the photon sphere, perihelion shift, gravitational redshift, Shapiro delay and deflection of light derived by this solution in Sec. \ref{sec:phenom}, and find bounds for the free parameters of the model preserving the standard predictions and the compatibility of GR. Finally, we present the main conclusions in Sec. \ref{sec:conclusions}. A general demonstration for the conservation law of the canonical energy-momentum tensor derived from the field equations of our metric-affine model is also presented in Appendix~\ref{sec:AppFieldEqs}.

We work with the metric signature $(+,-,-,-)$ and denote with a tilde on top quantities computed with respect to a generic affine connection, whereas their counterparts computed with respect to the Levi-Civita connection are denoted without a tilde. Latin and Greek indices run from $0$ to $3$, and refer to anholonomic and coordinate basis, respectively.

\section{Metric-Affine geometry}\label{sec:metricaffine}

The foundations of Riemannian geometry allow the introduction of smooth manifolds equipped with a metric tensor $g_{\mu\nu}$ and an affine connection $\Gamma^\lambda{}_{\mu\nu}$, which provide the notions of distance and parallel transport, respectively. Nevertheless, two fundamental assumptions are considered within this framework: the affine connection is both symmetric and metric-compatible. Accordingly, the Fundamental Theorem of Riemannian geometry leads to the uniqueness of such a connection, which is designated as the Levi-Civita connection and fully determined by the metric tensor:
\begin{equation}
\Gamma^{\lambda}\,_{\mu \nu}=\frac{1}{2}\,g^{\lambda \rho}\left(\partial_{\mu}g_{\nu \rho}+\partial_{\nu}g_{\mu \rho}-\partial_{\rho}g_{\mu \nu}\right)\,.
\end{equation}
Although this connection does not transform homogeneously under an infinitesimal coordinate transformation \newline $x^{\mu} \rightarrow x'^{\mu}=x^{\mu}+\xi^{\mu}$
\begin{equation}
\Gamma^{\lambda}\,_{\mu \nu} \rightarrow \Gamma '^{\lambda}\,_{\mu \nu}=\frac{\partial x^{\lambda}}{\partial x'^{\rho}}\frac{\partial x'^{\sigma}}{\partial x^{\mu}}\frac{\partial x'^{\omega}}{\partial x^{\nu}}\Gamma^{\rho}\,_{\sigma \omega}+\frac{\partial^{2} x'^{\rho}}{\partial x^{\mu}\partial x^{\nu}}\frac{\partial x^{\lambda}}{\partial x'^{\rho}}\,,
\end{equation}
it allows the definition of the intrinsic curvature of the manifold
\begin{equation}
[\nabla_{\mu},\nabla_{\nu}]\,v^{\lambda}=R^{\lambda}\,_{\rho \mu \nu}v^{\rho}\,,
\end{equation}
which suitably represents a tensor quantity depending on the metric tensor as follows:
\begin{equation}
R_{\lambda \rho \mu \nu}=\frac{1}{2}\left(\frac{\partial^{2}g_{\lambda \nu}}{\partial x^{\rho} \partial x^{\mu}}+\frac{\partial^{2}g_{\rho \mu}}{\partial x^{\lambda} \partial x^{\nu}}-\frac{\partial^{2}g_{\lambda \mu}}{\partial x^{\rho} \partial x^{\nu}}-\frac{\partial^{2}g_{\rho \nu}}{\partial x^{\lambda} \partial x^{\mu}}\right)+g_{\sigma \omega}\left(\Gamma^{\omega}\,_{\rho \mu}\Gamma^{\sigma}\,_{\lambda \nu}-\Gamma^{\omega}\,_{\rho \nu}\Gamma^{\sigma}\,_{\lambda \mu}\right)\,.
\end{equation}
By relaxing the previous assumptions, the geometry is extended towards an affinely connected metric space-time characterised by the additional post-Riemannian degrees of freedom present in the connection, namely the torsion and nonmetricity tensors
\begin{eqnarray}
T^{\lambda}\,_{\mu \nu}&=&2\tilde{\Gamma}^{\lambda}\,_{[\mu \nu]}\,,\\
Q_{\lambda \mu \nu}&=&\tilde{\nabla}_{\lambda}g_{\mu \nu}\,.
\end{eqnarray}

Under this geometrical prescription, torsion provides a displacement of vectors along infinitesimal paths that generally involves the breaking of standard parallelograms, whereas nonmetricity modifies lengths and angles under parallel transport \cite{Ortin:2015hya}. In addition, both quantities also fulfill a well-established correspondence with the continuous distribution of one-dimensional and point-like microstructure defects of crystal lattices \cite{Lobo:2014nwa}.

The decomposition of the affine connection separates each contribution into three different pieces, namely the Levi-Civita part and the so called contortion and disformation tensors depending on torsion and nonmetricity:
\begin{equation}\label{affineconnection}
\tilde{\Gamma}^{\lambda}\,_{\mu \nu}=\Gamma^{\lambda}\,_{\mu \nu}+K^{\lambda}\,_{\mu \nu}+L^{\lambda}\,_{\mu \nu}\,,
\end{equation}
with
\begin{eqnarray}
K^{\lambda}\,_{\mu \nu}&=&\frac{1}{2}\left(T^{\lambda}\,_{\mu \nu}-T_{\mu}\,^{\lambda}\,_{\nu}-T_{\nu}\,^{\lambda}\,_{\mu}\right)\,,\\
L^{\lambda}\,_{\mu \nu}&=&\frac{1}{2}\left(Q^{\lambda}\,_{\mu \nu}-Q_{\mu}\,^{\lambda}\,_{\nu}-Q_{\nu}\,^{\lambda}\,_{\mu}\right)\,.
\end{eqnarray}

This means a correction not only in the covariant derivative of sections of the vector bundle but also in the intrinsic curvature
\begin{equation}\label{totalcurvature}
\tilde{R}^{\lambda}\,_{\rho\mu\nu}=R^{\lambda}\,_{\rho\mu\nu}+\nabla_{\mu}N^{\lambda}\,_{\rho \nu}-\nabla_{\nu}N^{\lambda}\,_{\rho \mu}+N^{\lambda}\,_{\sigma \mu}N^{\sigma}\,_{\rho \nu}-N^{\lambda}\,_{\sigma \nu}N^{\sigma}\,_{\rho \mu}\,,
\end{equation}
where $N^{\lambda}\,_{\mu \nu}=K^{\lambda}\,_{\mu \nu}+L^{\lambda}\,_{\mu \nu}$ collects such corrections in the expression of the curvature tensor. Accordingly, the dynamics of the corresponding metric-affine fields can be considered by a gravitational action with higher order corrections, which also includes the respective contractions of the curvature tensor, namely the Ricci and co-Ricci tensors

\begin{eqnarray}\label{Riccitensor}
\tilde{R}_{\mu\nu}&=&\tilde{R}^{\lambda}\,_{\mu \lambda \nu}\,,\\
\label{co-Riccitensor}
\hat{R}_{\mu\nu}&=&\tilde{R}_{\mu}\,^{\lambda}\,_{\nu\lambda}\,,
\end{eqnarray}
as well as a homothetic curvature tensor $\tilde{R}^{\lambda}\,_{\lambda\mu\nu}$. In addition, the existence of a vector bundle isomorphism $e^{a}\,_{\mu}$ that allows the introduction of a local basis with its proper local metric structure at each point of the tangent bundle and of a principal bundle connection $\omega_{\mu} \in \mathfrak{gl}(4,R)$ compatible with the fundamental representations of spinor fields leads to a complete gauge characterisation of the metric-affine geometry in terms of the affine group $A(4,R) = R^{4} \rtimes GL(4,R)$ \cite{Hehl:1994ue}. Then the curvature, torsion and nonmetricity tensors acquire the following anholonomic expressions:
\begin{eqnarray}
F^{a}\,_{\mu\nu}&=&\partial_{\mu}e^{a}\,_{\nu}-\partial_{\nu}e^{a}
\,_{\mu}+\omega^{a}\,_{b\mu}\,e^{b}\,_{\nu}-\omega^{a}\,_{b\nu}\,e^{b}\,_{\mu}\,,
\\
F^{a}\,_{b\mu\nu}&=&\partial_{\mu}\omega^{a}\,_{b\nu}
-\partial_{\nu}\omega^{a}\,_{b\mu}+\omega^{a}\,_{c\mu}
\,\omega^{c}\,_{b}\,_{\nu}-\omega^{a}\,_{c\nu}
\,\omega^{c}\,_{b\mu}\,,\\
G_{ab\mu}&=&\partial_{\mu}g_{ab}-g_{ac}\,\omega^{c}\,_{b\mu}-g_{bc}\,\omega^{c}\,_{a\mu}\,,
\end{eqnarray}
where the coframe $e^{a}\,_{\mu}$ and the components $\omega^{a}\,_{b\mu}$ of the anholonomic connection satisfy the relations:
\begin{eqnarray}
g_{\mu \nu}&=&e^{a}\,_{\mu}\,e^{b}\,_{\nu}\,g_{a b}\,,\\
\label{anholonomic_connection}
\omega^{a}\,_{b\mu}&=&e^{a}\,_{\lambda}\,e_{b}\,^{\rho}\,\tilde{\Gamma}^{\lambda}\,_{\rho \mu}+e^{a}\,_{\lambda}\,\partial_{\mu}\,e_{b}\,^{\lambda}\,.
\end{eqnarray}

The structure of the $GL(4,R)$ group allows the definition of a large number of scalar invariants depending on the aforementioned quantities. In particular, the most general parity conserving quadratic Lagrangian includes $11 + 3 + 4$ irreducible modes from curvature, torsion and nonmetricity, respectively, which sets the search and analysis of exact solutions as an essential tool for the obtainment of a viable gravitational model in metric-affine geometry \cite{Hehl:1999sb}.

In order to find a physical configuration endowed with both dynamical torsion and nonmetricity fields, the following model based on Weyl-Cartan geometry (i.e. $Q_{\lambda\mu\nu}=g_{\mu\nu}W_{\lambda}$, with $W_{\lambda}$ the Weyl vector) has been recently considered in \cite{Bahamonde:2020fnq}:
\begin{eqnarray}\label{LagrangianIrreducible}
S &=& \int d^4x \sqrt{-g}
\left\{\mathcal{L}_{\rm m}+\frac{1}{64 \pi}\Bigl[
-\frac{4c^{4}}{G}R-6d_{1}\tilde{R}_{\lambda\left[\rho\mu\nu\right]}\tilde{R}^{\lambda\left[\rho\mu\nu\right]}-9d_{1}\tilde{R}_{\lambda\left[\rho\mu\nu\right]}\tilde{R}^{\mu\left[\lambda\nu\rho\right]}+8\,d_{1}\tilde{R}_{\left[\mu\nu\right]}\tilde{R}^{\left[\mu\nu\right]}
\Bigr.
\right.
\nonumber\\
& &
\left.
\Bigl.
\;\;\;\;\;\;\;\;\;\;\;\;\;\;\;\;\;\;\;+\,\frac{1}{8}\left(32e_{1}+8e_{2}+17d_{1}\right)\tilde{R}^{\lambda}\,_{\lambda\mu\nu}\tilde{R}^{\rho}\,_{\rho}\,^{\mu\nu}-7d_{1}\tilde{R}_{\left[\mu\nu\right]}\tilde{R}^{\lambda}\,_{\lambda}\,^{\mu\nu}+3\left(1-2a_{2}\right)T_{\left[\lambda\mu\nu\right]}T^{\left[\lambda\mu\nu\right]}\Bigr]\right\}\,.
\end{eqnarray}

In the context of the gauge formalism, the following field equations are obtained by performing variations with respect to the coframe $e^{a}\,_{\mu}$ and the spin connection $\omega^{a}\,_{b\mu}$:
\begin{eqnarray}\label{field_eq1}
X1_{\mu}\,^{\nu} &=& 16\pi\theta_{\mu}\,^{\nu}\,,\\
\label{field_eq2}
X2^{\lambda\mu\nu} &=& 16\pi\bigtriangleup^{\lambda\mu\nu}\,,
\end{eqnarray}
where $X1_{\mu}\,^{\nu}$ and $X2^{\lambda\mu\nu}$ are tensor quantities defined in Appendix \ref{sec:AppFieldEqs}, whereas $\theta_{\mu}\,^{\nu}$ and $\bigtriangleup^{\lambda\mu\nu}$ describe the canonical energy-momentum and hypermomentum density tensors of matter, respectively:
\begin{eqnarray}
\theta_{\mu}\,^{\nu}&=&\frac{e^{a}\,_{\mu}}{\sqrt{- g}}\frac{\delta\left(\mathcal{L}_{\rm m}\sqrt{- g}\right)}{\delta e^{a}\,_{\nu}}\,,\\
\bigtriangleup^{\lambda\mu\nu}&=&\frac{e^{a\lambda}e_{b}\,^{\mu}}{\sqrt{- g}}\frac{\delta\left(\mathcal{L}_{\rm m}\sqrt{-g}\right)}{\delta\omega^{a}\,_{b\nu}}\,.
\end{eqnarray}

Therefore, both matter currents act as sources of the extended gravitational field. In particular, the general decomposition of the affine connection (\ref{affineconnection}) and, correspondingly, of the anholonomic connection (\ref{anholonomic_connection}) leads hypermomentum to present its proper decomposition into spin and dilation currents, which in the realm of Weyl-Cartan geometry induces space-time torsion and nonmetricity according to the internal structure of the homogeneous Weyl group $W(1,3) \subset GL(4,R)$ \cite{Kopczynski:1988jq}.

As can be seen from the action (\ref{LagrangianIrreducible}), the field strength tensors of torsion and nonmetricity represent dynamical deviations from the Bianchi identities of GR and their contractions
\begin{equation}\label{curvbianchi}
\tilde{R}^{\lambda}\,_{[\mu \nu \rho]}+\tilde{\nabla}_{[\mu}T^{\lambda}\,_{\nu \rho]}+T^{\sigma}\,_{[\mu \nu}\,T^{\lambda}\,_{\rho] \sigma}=0\,,\end{equation}
\begin{equation}\label{riccibianchi}
\tilde{R}_{[\mu \nu]}=\frac{1}{2}\tilde{R}^\lambda\,_{\lambda\mu\nu}+\tilde{\nabla}_{[\mu}T^{\lambda}\,_{\nu]\lambda}+\frac{1}{2}\tilde{\nabla}_{\lambda}T^{\lambda}\,_{\mu\nu}-\frac{1}{2}T^{\lambda}\,_{\rho\lambda}T^{\rho}\,_{\mu\nu}\,,
\end{equation}
\begin{equation}\label{nonmetricitybianchi}
\tilde{R}^{\lambda}\,_{\lambda\mu\nu}=4\nabla_{[\nu}W_{\mu]}\,.
\end{equation}

Furthermore, by taking the trace of Eq. (\ref{field_eq1}), it is straightforward to check that the Riemannian scalar curvature must vanish, which represents a strong geometrical constraint involving the metric tensor alone and significantly restricts the Riemannian part of the space-time geometry within this model. On the other hand, the respective trace of Eq. (\ref{field_eq2}) allows us to describe the weak-field limit of torsion and nonmetricity as follows:
\begin{eqnarray}\label{weaktorsion}
\nabla_{\rho}\nabla_{\lambda}T^{\lambda\rho}\,_{\mu}+\nabla_{\rho}\nabla^{\rho}T^{\lambda}\,_{\mu\lambda}-\nabla_{\rho}\nabla_{\mu}T^{\lambda\rho}\,_{\lambda}=0\,,\end{eqnarray}
\begin{eqnarray}
\label{weaknonmetricity}
\nabla_{\mu}\tilde{R}^{\lambda}\,_{\lambda}\,^{\mu\nu}=0\,.\end{eqnarray}

The absence of the Birkhoff’s theorem for the general quadratic gravitational action of MAG allows the existence of nontrivial spherically symmetric vacuum solutions beyond the Schwarzschild geometry in the presence of torsion and nonmetricity \cite{Neville:1979fk,Rauch:1981tva,ho1997some} (see \cite{Obukhov:2020hlp} for a natural extension with parity violating terms in the torsion sector). Therefore, the dynamical effects of these quantities can potentially be yielded by a gravitational configuration with spin and dilation charges, which obeys the field equations (\ref{field_eq1}) and (\ref{field_eq2}) with $\theta_{\mu}\,^{\nu}=\bigtriangleup^{\lambda\mu\nu}=0$, and shows the following tensor components \cite{Hohmann:2019fvf}:
\begin{eqnarray}
T^{t}\,_{t r}&=&a(r) \,,\\
T^{r}\,_{t r}&=&b(r)\,,\\
T^{\theta_{k}}\,_{t \theta_{k}}&=&f(r)\,,\\
T^{\theta_{k}}\,_{r \theta_{k}}&=&g(r) \,,\\
T^{\theta_{k}}\,_{t \theta_{l}}&=&e^{a \theta_{k}}\,e^{b}\,_{\theta_{l}}\,\epsilon_{a b}\, d (r) \,,\\
T^{\theta_{k}}\,_{r \theta_{l}}&=&e^{a \theta_{k}}\,e^{b}\,_{\theta_{l}}\,\epsilon_{a b}\, h (r) \,,\\
T^{t}\,_{\theta_{k} \theta_{l}}&=&\epsilon_{k l} \, k (r)\,\sin\theta_1 \,,\\
T^{r}\,_{\theta_{k} \theta_{l}}&=&\epsilon_{k l} \, l (r)\,\sin\theta_1 \,,\\
W_\lambda&=&\left(w_{1}(r),w_{2}(r),0,0\right)\,,
\end{eqnarray}
\begin{equation}\label{metric}
    ds^2=c^{2}\Psi_{1}(r)\,dt^2-\frac{dr^2}{\Psi_{2}(r)}-r^2\left(d\theta_1^2+\sin\theta_1^2d\theta_2^2\right)\,,
\end{equation}
where $\epsilon_{kl}$ is the Levi-Civita symbol in two dimensions. The requirement of regularity restricts the initial arbitrariness of the torsion components by imposing the relations
\begin{eqnarray}\label{rel1}
b(r)&=&c\,a(r)\,\sqrt{\Psi_{1}(r)\Psi_{2}(r)}\,,\;\;\;\;\;\;f(r) = - \,c\, g(r)\,\sqrt{\Psi_{1}(r)\Psi_{2}(r)}\,,
\;\;\;\;\;\;\;\;\;\;\;\;
\nonumber\\
d(r)&=&- \,c\,h(r)\,\sqrt{\Psi_{1}(r)\Psi_{2}(r)}\,,\;\;\;l(r) = c\,k(r)\,\sqrt{\Psi_{1}(r)\Psi_{2}(r)}\,,
\end{eqnarray}
and of the Weyl vector components
\begin{equation}\label{rel2}
w_{1}(r)=-\,c\,w_{2}(r)\sqrt{\Psi_{1}(r)\Psi_{2}(r)}\,.
\end{equation}
Then, it can be easily shown that the rest of the free functions related to the spherically symmetric case are completely determined by the field equations and the weak-field limit. In this regard, a straightforward integration of the expressions (\ref{weaktorsion}) and (\ref{weaknonmetricity}) gives rise to the following solutions:
\begin{eqnarray}
w_{1}(r)&=&-\,c\,\kappa_{d}\,\int\sqrt{\frac{\Psi_{1}(r)}{\Psi_{2}(r)}}\,\frac{dr}{r^2}\,, \label{rel3}\\
b(r)&=&rf\,'(r)+f(r)+\frac{c\,\kappa_{d}}{2r}\,
\sqrt{\frac{\Psi_{1}(r)}{\Psi_{2}(r)}}\,,\label{rel4}
\end{eqnarray}
where we use the prime symbol to denote differentiation with respect to the radial coordinate. In addition, by considering the compatibility with Coulomb electric and magnetic charges, it is clear that our final solution must fulfill the condition $\Psi_{1}(r)=\Psi_{2}(r)$ and one may assume such a condition for simplicity. In any case, the same result turns out to be obtained by the field equations of the model, as shown below.

In the first place, by taking into account the relations (\ref{rel1})-(\ref{rel4}), the symmetric component of the field equation (\ref{field_eq2}) is reduced to the constraint $e_{2}=-\,d_{1}/2$. This relation sets the coefficient $e_{1}$ as the unique coupling constant describing the dynamics of nonmetricity within this model. On the other hand, the antisymmetric components $X2^{[01]0}$, $X2^{[02]2}$ and $X1^{[01]}$ provide the following system of independent equations:
\begin{eqnarray}
0&=&k\left(r\Psi_{1}\Psi'_{2}h+r\Psi'_{1}\Psi_{2}h+2r\Psi_{1}\Psi_{2}h'+6h \Psi_{1}\Psi_{2}\right)\,,\\
0&=&4c^2\Psi_{1}\Psi_{2}k\left(k\Psi_{1}\right)'-2r^2\Psi_{2}\left(r^2h^2\Psi_{1}\Psi_{2}\right)'-4c\,r^2\Psi_{1}^{3/2}\Psi_{2}^{3/2}k'h-6c\,r^{2}\Psi_{1}^{3/2}\Psi_{2}^{3/2}kh'
\nonumber\\
&&-3c\,r^{2}\Psi_{1}^{3/2}\Psi_{2}^{1/2}\Psi_{2}'kh-7c\,r^{2}\Psi_{1}^{1/2}\Psi_{2}^{3/2}\Psi_{1}'kh-2c\,r\Psi_{1}^{3/2}\Psi_{2}^{3/2}kh\,,\\
0&=&2c\,r^2\Psi_{1}^{3/2}\Psi_{2}^{3/2}\Psi'_{1}kh+2c^2\Psi_{1}^{2}\Psi_{2}\Psi'_{1}k^{2}-4r^3\Psi_{1}^{2}\Psi_{2}^{2}h^2-4c\,r\Psi_{1}^{5/2}\Psi_{2}^{3/2}kh+2c\,r^2\Psi_{1}^{5/2}\Psi_{2}^{3/2}k'h
\nonumber\\
&&-2r^4\Psi_{1}\Psi_{2}^{2}\Psi'_{1}h^2-2r^4\Psi_{1}^{2}\Psi_{2}\Psi'_{2}h^2-4r^4\Psi_{1}^{2}\Psi_{2}^{2}hh'+c\,r^3\Psi_{1}^{3/2}\Psi_{2}^{1/2}\Psi'_{1}\Psi'_{2}kh+c\,r^3\Psi_{1}^{5/2}\Psi_{2}^{1/2}\Psi'_{2}k'h
\nonumber\\
&&+2c\,r^3\Psi_{1}^{3/2}\Psi_{2}^{3/2}\Psi'_{1}kh'+c\,r^3\Psi_{1}^{3/2}\Psi_{2}^{3/2}\Psi'_{1}k'h+2c^2\Psi_{1}^{3}\Psi_{2}kk'+2c\,r^3\Psi_{1}^{5/2}\Psi_{2}^{3/2}k'h'+c\,r^3\Psi_{1}^{1/2}\Psi_{2}^{3/2}\Psi'{}_{1}^{2}kh\,.
\end{eqnarray}

These equations have the following nonvanishing solutions:
\begin{equation}\label{dyntorsionfunction}
h(r)=-\,\kappa_{s}/(r\sqrt{\Psi_{1}(r)\Psi_{2}(r)})\,,\quad k(r)=0\,,
\end{equation}
\begin{equation}
k(r)=\beta/\Psi_{1}(r)\,,\quad h(r)=0\,,
\end{equation}
which describe two different branches with integration constants $\kappa_{s}$ and $\beta$. Nevertheless, only the first branch (\ref{dyntorsionfunction}) provides a dynamical contribution for the torsion tensor. As can be seen, for the second case the combination $2c^2\beta\sin\theta_{1} X2^{[23]0}+X1_{0}\,^{0}-X1_{1}\,^{1}=0$ is reduced to the following condition:
\begin{equation}
\left(\ln{\frac{\Psi_{1}}{\Psi_{2}}}\right)'=0\,.
\end{equation}
Such condition involves the proportionality relation $\Psi_{1}=\alpha\Psi_{2} \equiv \Psi(r)$ between the metric functions, where we can set $\alpha=1$ without any loss of generality by a redefinition of the time coordinate. This equality establishes the Reissner-Nordstr\"{o}m (RN) metric as the only possible static and spherically symmetric vacuum solution within this model, in virtue of the vanishment of the Riemannian scalar curvature derived from the trace of the Expression (\ref{field_eq1}). In addition, the antisymmetric component $X1^{[23]}$ of the field equations is vanished either for $a_{2}=1/2$ or $\beta=0$. The latter switches off any independent contribution of the torsion tensor on the field strength tensors $\tilde{R}^{\lambda}\,_{[\mu \nu \rho]}$ and $\tilde{R}_{[\mu \nu]}$, which corresponds to the absence of dynamical torsion in that case. The same conclusion is reached if $\beta \neq 0$ and $a_{2}=1/2$ when imposing the field equation $X2^{[23]0}=0$, which sets the branch (\ref{dyntorsionfunction}) as the unique one providing a dynamical torsion contribution.

For this branch, the combination $c\,\sqrt{\Psi_{1}\Psi_{2}}\,X2^{[23]0}-X2^{[23]1}=0$ leads to the corresponding equality of the metric functions $\Psi_{1}=\Psi_{2}$ and the remaining field equations described by the Expression (\ref{field_eq2}) provide the following algebraic expression for the function $g$:
\begin{equation}
g(r)=-\,\frac{1}{2r}-\frac{wr}{2\Psi(r)}-\frac{\kappa_{d}}{2r\Psi(r)}\,,
\end{equation}
with $w=\left(1-2a_{2}\right)/d_{1}$. Finally, the Expression (\ref{field_eq1}) is reduced to the following independent equation depending on the metric function
\begin{equation}
r^2-r^2\Psi(r)-r^3\Psi'(r) = \frac{G\left(d_{1}\kappa_{s}^{2}-4e_{1}\kappa_{d}^{2}\right)}{c^4}\,,
\end{equation}
which has as solution, the RN metric
\begin{eqnarray}\label{RN}
\Psi(r)&=&1-\frac{2GM}{c^{2}r}+\frac{G\left(d_{1}\kappa^{2}_{s}-4e_{1}\kappa^{2}_{d}\right)}{c^{4}r^2}\,.
\end{eqnarray}

It is worthwhile to stress the decomposition of the field strength tensors
\begin{eqnarray}
\tilde{R}^{\lambda}\,_{[\mu\nu\rho]}&=&\frac{1}{4}\tilde{R}^{\sigma}\,_{\sigma[\mu\nu}\delta_{\rho]}\,^{\lambda}+\bar{R}^{\lambda}\,_{[\mu\nu\rho]}\,,\\
\tilde{R}_{[\mu\nu]}&=&\frac{1}{4}\tilde{R}^{\sigma}\,_{\sigma\mu\nu}+\bar{R}_{[\mu\nu]}\,,
\end{eqnarray}
as a sum of the homothetic curvature $\tilde{R}^{\sigma}{}_{\sigma\mu\nu}=4\nabla_{[\nu}W_{\mu]}$ and the effective dynamical strength contribution of torsion in the solution. In this case, the resulting gravitational action expressed in terms of this decomposition acquires the following simple form:
\begin{eqnarray}
S &=& \frac{1}{64 \pi}\int d^4x \sqrt{-g}
\Bigl[
-\frac{4c^{4}}{G}R-6d_{1}\bar{R}_{\lambda\left[\rho\mu\nu\right]}\bar{R}^{\lambda\left[\rho\mu\nu\right]}-9d_{1}\bar{R}_{\lambda\left[\rho\mu\nu\right]}\bar{R}^{\mu\left[\lambda\nu\rho\right]}+8\,d_{1}\bar{R}_{\left[\mu\nu\right]}\bar{R}^{\left[\mu\nu\right]}
\Bigr.
\nonumber\\
& &
\Bigl.
\;\;\;\;\;\;\;\;\;\;\;\;\;\;\;\;\;\;\;\;\;\;\;\;\;\;+\,4e_{1}\tilde{R}^{\lambda}\,_{\lambda\mu\nu}\tilde{R}^{\rho}\,_{\rho}\,^{\mu\nu}+3\left(1-2a_{2}\right)T_{\left[\lambda\mu\nu\right]}T^{\left[\lambda\mu\nu\right]}\Bigr]\,.
\end{eqnarray}

The nature of the horizons provided by the solution (\ref{RN}) depends on the corresponding balance of the spin and dilation charges $\kappa_{s}$ and $\kappa_{d}$, through the difference $d_{1}\kappa^{2}_{s}-4e_{1}\kappa^{2}_{d}$. Thus, a positive difference of this quantity would present two horizons determined from the roots
\begin{eqnarray}
r_{\pm}=\frac{G}{c^2}\left(M\pm \Delta_1\right)\,,\quad \Delta_1^2= M^2-\frac{1}{G}\left(d_1\kappa_s^2-4e_1\kappa_d^2\right)\,,\label{Delta}
\end{eqnarray}
with $0 < \frac{1}{G}\left(d_1\kappa_s^2-4e_1\kappa_d^2\right) < M^2$. Nevertheless, the different signs of the kinetic terms related to the dynamical part of torsion and to the Weyl vector allows the case $d_{1}\kappa^{2}_{s}-4e_{1}\kappa^{2}_{d} < 0$, which is characterised by the absence of an inner Cauchy horizon and sets a unique event horizon at the root $r_{+}$. A similar situation occurs in the extreme case $M^2=\frac{1}{G}\left(d_1\kappa_s^2-4e_1\kappa_d^2\right)$ where both horizons coincide and furthermore if $M^2 < \frac{1}{G}\left(d_1\kappa_s^2-4e_1\kappa_d^2\right)$, where the solution does not contain any horizon but a naked singularity. In this sense, the balance between the spin and dilation charges is not restricted to any special constraint and therefore any of these situations may occur in the presence of torsion and nonmetricity, in contrast to the standard RN counterpart with electromagnetic fields of GR.

\section{Observational constraints}\label{sec:phenom}
In the general metric-affine regime, the existing correspondence between the hypermomentum density tensor and the post-Riemannian degrees of freedom contained in the linear connection provide a generalised conservation law for the canonical energy-momentum tensor of matter \cite{Hehl:1994ue} (see Appendix \ref{sec:AppFieldEqs} for its derivation in the present model of MAG):
\begin{equation}\label{conslaw}
\nabla_{\nu}\theta_{\mu}\,^{\nu}+\left(K_{\lambda\rho\mu}+L_{[\lambda\rho]\mu}\right)\theta^{\left[\rho\lambda\right]}+\tilde{R}_{\lambda\rho\sigma\mu}\bigtriangleup^{\rho\lambda\sigma}=0\,.
\end{equation}

Likewise, the latter also provides the equations of motion of test bodies with microstructure coupled to the torsion and nonmetricity tensors by the definition of their multipole moments \cite{Puetzfeld:2007hr}. Indeed, by integrating the Expression (\ref{conslaw}) over a three dimensional space-like section of a world tube involving the test body, and expanding the affine connection and the curvature tensor at first order
\begin{equation}
\frac{d}{dt}\int{\theta^{\mu t}\sqrt{-g}\,d^{3}x'}+\Gamma^{\mu}\,_{\lambda\rho}\int{\theta^{\lambda \rho}\sqrt{-g}\,d^{3}x'}+N_{[\lambda\rho]}\,^{\mu}\int{\theta^{[\rho\lambda]}\sqrt{-g}\,d^{3}x'}+\tilde{R}_{\lambda\rho\sigma}\,^{\mu}\int{\bigtriangleup^{\rho\lambda\sigma}\sqrt{-g}\,d^{3}x'}=0\,,
\end{equation}
it is possible to identify the resulting integrals with the four-momentum, velocity and hypermomentum tensor of the test body as follows,
\begin{eqnarray}
p^{\lambda}u^{\rho}&=&\dot{t}\int{\theta^{\lambda\rho}\sqrt{-g}\,d^{3}x'}\,,\\
\bigtriangleup^{\lambda\rho}u^{\sigma}&=&\dot{t}\int{\bigtriangleup^{\lambda\rho\sigma}\sqrt{-g}\,d^{3}x'}\,,
\end{eqnarray}
where the dot denotes differentiation with respect to an affine parameter $\lambda$. Accordingly, the following equations of motion in MAG are straightforwardly obtained:
\begin{equation}\label{eqsmotion}
\dot{p}^{\mu}+\Gamma^{\mu}\,_{\lambda \rho}\,p^{\lambda}u^{\rho}+N_{[\lambda \rho]}\,^{\mu}p^{\rho}u^{\lambda}+\tilde{R}_{\lambda \rho \sigma}\,^{\mu}\bigtriangleup^{\rho\lambda}u^{\sigma}=0\,.
\end{equation}

As can be seen, these equations present additional Lorentz-like forces depending on the intrinsic hypermomentum of matter, which also determines the antisymmetric part of its canonical energy-momentum tensor, in such a way that they can only operate on the trajectories of spinning and dilational bodies. Nevertheless, for the rest of ordinary matter uncoupled to the torsion and nonmetricity tensors, the trajectories describe a geodesic motion which can also contain deviations induced by the dynamical contribution of these tensors in the Levi-Civita connection. Therefore, the existence of a fully relativistic configuration with both dynamical torsion and nonmetricity fields as the one considered by our model allows us to study these deviations in the framework of MAG.

For this task, the geodesic part of Eq. (\ref{eqsmotion}) turns out to be suppressed in the equatorial plane $\theta_{1}=\pi/2$ and described by the following system of equations \cite{Weinberg}
\begin{eqnarray}
\label{geo1}
\ddot{t}&=&-\,\left(\frac{\Psi'(r)}{\Psi(r)}\right)\dot{t}\,\dot{r}\,,\\
\label{geo2}
\ddot{r}&=&r\Psi(r)\,\dot{\theta}_{2}^{2}-\,\frac{\Psi'(r)}{2\Psi(r)}\left(c^{2}\Psi^{2}(r)\,\dot{t}^{2}-\,\dot{r}^{2}\right)\,,\\
\label{geo3}
\ddot{\theta}_{2}&=&-\,\left(\frac{2}{r}\right)\,\dot{r}\dot{\theta}_{2}\,.
\end{eqnarray}

From (\ref{geo1}) and (\ref{geo3}), one finds the following constants of motion, related to the energy of the test body and the orbital angular momentum 
\begin{eqnarray}\label{kh}
E &:=&\left(1-\frac{2GM}{c^2r}+\frac{G(d_{1}\kappa^{2}_{s}-4e_{1}\kappa^{2}_{d})}{r^2c^4}\right)\dot{t}\,,\\
	J &:=&r^2 \dot{\theta}_{2}\,.
\end{eqnarray}
Furthermore, if we introduce the above conserved quantities in (\ref{geo2}) we find
\begin{align}\label{17}
\dot{r}^2=c^2E^2-\frac{\Psi(r)J^2}{r^2}-c^2\sigma\Psi(r)\,,
\end{align}
where $\sigma=0 (\sigma=1)$ represents  massless(massive)  particles. Then, the above equation can be rewritten in terms of an effective potential, namely
\begin{align}
	\frac{1}{2}\dot{r}^2 + V(r)=0\,,\quad V(r)=-\,\frac{1}{2} c^2E^2 + \frac{1}{2} \Psi(r) \left( \frac{J^2}{r^2} + \sigma c^2  \right)\,,\label{potentialEq}
\end{align}
or reparametrised by $r(\lambda) \rightarrow r(\theta_{2})$, yielding
\begin{align}\label{eq:rphi}
	\frac{1}{2}\frac{\dot r^2}{\dot{\theta}_{2}^{2}} + \frac{1}{\dot{\theta}_{2}^{2}}V(r)  = \frac{1}{2}\left(\frac{dr}{d\theta_{2}}\right)^2 + \frac{r^4}{J^2}V(r)=0\,.
\end{align}
Now, we have all the ingredients needed to find different observables related to our spherically symmetric solution.

\subsection{Photon sphere and perihelion shift}
The photon sphere of the solution~\eqref{RN} can be obtained by considering the roots ($r,J$) of the effective potential and its first derivative for circular photon orbits (i.e. $V'(r)=V(r)=\sigma=0$) \cite{Claudel:2000yi}. Thereby, considering the effective potential defined in \eqref{potentialEq}, one then finds the following roots
\begin{eqnarray}
r_{1}&=&\frac{G}{2c^2} \left(3 M+ \Delta_2\right)\,,\quad J_{1,\pm}=
\pm\frac{G E (\Delta_2+3 M)^2}{\sqrt{2} c \sqrt{\Delta_2^2+3 M^2+4 \Delta_2 M}}\,,\\
r_2&=&\frac{G}{2c^2} \left(3 M- \Delta_2\right)\,,\quad J_{2,\pm}=\pm\frac{G E (\Delta_2-3 M)^2}{\sqrt{2} c \sqrt{\Delta_2^2+3 M^2-4 \Delta_2 M}}\,,\label{shadow2}
\end{eqnarray}
where we have defined 
\begin{eqnarray}
\Delta_2^2:&=&M^2+8\Delta_1^2=9M^2-\frac{8}{G} \left(d_1 \kappa_s^2-4 e_1 \kappa_d^2\right)\,.
\end{eqnarray}

Note that the roots $(r_{2},J_{2,\pm})$ are not well-defined for the black hole case $0 \leq \frac{1}{G}\left(d_1\kappa_s^2-4e_1\kappa_d^2\right) \leq M^2$, and accordingly they cannot represent a photon sphere. Conversely, the first pair $(r_{1},J_{1,\pm})$ describes a unique photon sphere that lies outside the event horizons, with the corrections related to the spin and dilation charges affecting its location in the equatorial plane with respect to the Schwarzschild solution of GR.

It is well-known that the photon sphere located at $r_{+}$ creates a certain patch in the sky around the black hole that cannot be observed. This is known as the shadow of the black hole from which the light from distant sources cannot reach the observer. For our metric, the shadow of the black hole presents a RN type but its size can be affected due to the fact that the new terms depending on the hypermomentum charges may provide an effective negative contribution. In any case, the size of the shadow cannot be used to constrain the model from the current data of M87* in a realistic way, since for this astrophysical object one would need to consider an axisymmetric metric to get corrections also in its shape~\cite{Akiyama:2019cqa}.

On the other hand, in order to derive the perihelion shift provided by our solution, we consider a massive body with $\sigma=1$ and a perturbation around its closed orbit $r_c$ (i.e. $\dot{r}_c=V(r_c)=V'(r_c)=0$) so that the radial coordinate in \eqref{eq:rphi} can be expressed as $r(\phi)= r_{c} + r_{\phi}(\phi)$. For simplicity, we have also renamed the angle $\theta_{2}$ as $\phi$. Then, one can assume that the perturbation $r_\phi/r_c\ll 1$ is small. By doing the expansion around $r_\phi/r_c$, one finds that \eqref{eq:rphi} becomes
\begin{align}
    \left(\frac{d r_\phi}{d\phi}\right)^2 = -\,2 \frac{(r_c + r_\phi)^4}{J^2} V(r_c + r_\phi)= -\,\frac{r_c^4}{J^2} V''(r_c)r_{\phi}^{2} + \mathcal{O}\left(\tfrac{r_{\phi}^{3}}{r_c^{3}}\right) \,,
\end{align}
where the angular momentum constant $J$ reads
\begin{eqnarray}
J_{\pm}=\pm\, r \sqrt{\frac{G c^2(c^2 M r-d_1 \kappa_s^2+4 e_1 \kappa_d^2)}{c^4 r^2-3 c^2 G M r+2 d_1 G \kappa_s^2-8e_1G\kappa_d^2}}\,.\label{h+-}
\end{eqnarray}

The solution of the above equation for $r(\phi)$ oscillates with a wave number $K = \sqrt{\frac{r_c^4}{J^2} V''(r_c)} $, and then the perihelion shift becomes
\begin{align}
    \Delta \phi =2\pi\left(\frac{1}{K}-1\right)\,.
\end{align}
Then, by replacing the values of $J$ 
and $V''(r_c)$ in the massive case expressed in~\eqref{h+-} and \eqref{potentialEq} for our RN type of solution, the general expression of the perihelion shift acquires the form
\begin{eqnarray}
\Delta \phi=2 \pi  \left(c^2 r_c \sqrt{\frac{c^2 M r_c-d_1 \kappa_s^2+4 e_1 \kappa_d^2}{c^6 M r_c^3-6 c^4 G M^2 r_c^2+9 c^2 G M r_c \left(d_1 \kappa_s^2-4 e_1 \kappa_d^2\right)-4 G \left(d_1 \kappa_s^2-4 e_1 \kappa_d^2\right)^2}}-1\right)\,.
\end{eqnarray}
Since $r_c\gg 2GM/c^2$ and $r_c^2 \gg G(d_1\kappa_s^2-4e_1 \kappa_d^2)/c^4$, we can expand the above expression up to fourth order yielding
\begin{eqnarray}\label{DeltaphiF}
  \Delta \phi=2\pi\bigg[\frac{3 GM}{c^2r_c}+\frac{27 G^2M^2 }{2 c^4r_c^2}+\frac{135G^3 M^3 }{2 c^6r_c^3}+\frac{2835 G^4M^4 }{8c^8 r_c^4}-d_1 \kappa_s^2 \left(\frac{1}{2c^2 M r_c}+\frac{6G }{c^4r_c^2}\right)+e_1 \kappa_d^2 \left(\frac{2 }{M c^2r_c}+\frac{24G}{c^4r_c^2}\right)\bigg]\,.
\end{eqnarray}
For an elliptical orbit, we have $r_c=a(1-e^2)$ where $a$ is the semi-axis and $e$ the eccentricity of the orbit. The first four terms describe the standard Schwarzschild contributions up to fourth order expansion whereas the rest corresponds to the new contributions yielded by the spin and dilation charges.

One could expect that these contributions coming from metric-affine geometry will be only sourced in a strong gravitational regime. One possible example of this could occur in very dense objects such as a supermassive black hole. Thus, the high precision astrometric observations close to the compact source Sgr A* and particularly of the S2 star orbiting the supermassive black hole allow us to estimate such contributions \cite{Abuter:2020dou}. In the following, whenever we use numerical values, we will take the gravitational constant as $G=6.6743\cdot 10^{-11}\,\textrm{m}^3/(\textrm{kg}\,\textrm{s}^2)$ and the velocity of light as $c=299,792,458\, \textrm{m}/\textrm{s}$. For the S2 start orbiting Sgr A*, the mass of the black hole is $M_{\rm Sgr A*}=4.260 \cdot 10^{6}M_{\odot}$ with $M_{\odot}= 1.989\cdot 10^{30} \,\textrm{kg}$  and since $M_{\rm S_2}$ has around $10-15M_{\odot}$, one can approximate the orbit of the S2 star as a test body following a geodesic motion. The period of the orbit for S2 has been measured to be around $16.052\, \textrm{years}$ and since it is not totally circular, one can replace $r_c$ by $a(1-e^2)$. For our case, $a_{\rm S_2}=970\,\textrm{au}$ and $e_{\rm \rm S_2}=0.884649$, which gives us that the quantity $r_c = a(1-e^2) = 3.142\cdot 10^{13} \, \textrm{m}$~\cite{Abuter:2020dou}. For the Schwarzschild contribution coming from the first four terms in~\eqref{DeltaphiF}, we get that the perihelion shift predicted for this system is
\begin{equation}
    \Delta \phi_{\rm  S_2}^{(\textrm{GR})}\approx 
     48.550\,\left[^{''}/\textrm{year}\right]\,.
    \end{equation}
On the other hand, the observed value of the perihelion shift for the S2 star around Sgr A* is~\cite{Abuter:2020dou}
\begin{align}
 \Delta \phi_{\rm  S_2}^{(\textrm{obs})}=48.506 f_{\rm SP}\,\left[^{''}/\textrm{year}\right]\,,
\end{align}
where $f_{\rm SP}$ is a  fitting parameter that for this system, it has its best observational value between $f_{\rm SP,\rm max}=1.2$ and $f_{\rm SP,\rm min}=0.9$. Then, the difference between the GR predicted value and the observed maximum and minimum values are
\begin{align}
  \Big( \Delta \phi^{(\textrm{obs})}_{\rm max}- \Delta\phi^{(\textrm{GR})}\Big)_{\rm  S_2} &= 9.657\,\left[^{''}/\textrm{year}\right]\,,\\
  \Big( \Delta \phi^{(\textrm{obs})}_{\rm min}- \Delta\phi^{(\textrm{GR})}\Big)_{\rm S_2} &= -\,4.894\,\left[^{''}/\textrm{year}\right]\,.
\end{align}
Therefore, we can now constrain the parameters coming from the modification of GR that are related to $d_1$ and $e_1$. Using~\eqref{DeltaphiF} for the complete calculation of the perihelion shift of the S2 star, we can then find that the value of $d_1\kappa_s^2$ must lie between
\begin{eqnarray}\label{Sagi}
4e_1\kappa_d^2- 5.711\cdot 10^{63} \, \left[J\cdot m\right]\leq d_{1}\kappa_s^2\leq 4e_1\kappa_d^2+
2.894\cdot 10^{63} \, \left[J\cdot m\right]\,.
\end{eqnarray}

It is clear that the current observations allow the case with a negative factor $d_{1}\kappa^{2}_{s}-4e_{1}\kappa^{2}_{d}$ in the RN solution, where the corresponding inner horizon is absent and therefore the breakdown of predictability and the presence of other instabilities may be avoided \cite{Poisson:1990eh,Maeda:2005yd}.

\subsection{Gravitational redshift}
Another gravitational effect that can be used to constrain the new effects arising from our model is the gravitational redshift. It is well known that the wavelength or frequency of electromagnetic waves or photons is modified  when they passed near an object due to its gravitational field. The photons lose energy, so that, their wavelength is increased and their frequency is decreased, which creates a redshift. Let us suppose that a ray of light is propagating at different heights $r_1$ and $r_2$ (with $r_1<r_2$) in a  gravitational field which is characterised by the metric~\eqref{metric}. The gravitational redshift $z$ can be defined as a measure depending on the quotient between the frequencies of the photon $\nu_1$ and $\nu_2$, which can be written in terms of the metric tensor as
\begin{eqnarray}
z=\frac{\nu_2}{\nu_1}-1=\sqrt{\frac{\Psi(r_2)}{\Psi(r_1)}}-1\,.
\end{eqnarray}
Let us now assume that $\nu_1$ is the frequency measured at the source of emission, so that for an astrophysical system 
we can assume that $r_1=R$, where $R$ is the radius of the source. Then, $\nu_2$ is the frequency measured by an observer at infinity $r_2\rightarrow \infty$. For astrophysical systems, this condition holds since we receive the electromagnetic signal from an object which is far away from the source. Now, we take the solution~\eqref{RN},  and assume the gravitational potential to be small $\Psi\approx 1$, which means that
\begin{eqnarray}
 r\gg \frac{2GM}{c^2}\,,\quad r^2\gg \frac{G(d_1\kappa_s^2-4e_1 \kappa_d^2)}{c^4}\,.\label{approx}
\end{eqnarray}
These approximations can be easily computed if one introduces a small tracking parameter $\epsilon\ll 1$ such that one re-scales the parameters as $M\rightarrow \epsilon M,\, \kappa_d\rightarrow \epsilon^2\kappa_d$ and $ \kappa_s\rightarrow \epsilon^2\kappa_s$. By doing this, we find that the gravitational redshift up to fourth order in $\epsilon$ becomes
\begin{eqnarray}
z=\frac{G M}{c^2 R}+\frac{3 G^2 M^2}{2 c^4 R^2}+\frac{5 G^3 M^3}{2 c^6 R^3}+\frac{35 G^4 M^4}{8 c^8 R^4}+\frac{G \left(4 e_1 \kappa_d^2-d_1 \kappa_s^2\right)}{2 c^4 R^2}\,.\label{redshift}
\end{eqnarray}

We can now consider astrophysical compact objects, such as degenerate stars composed by fermionic matter with a spin alignment induced by torsion, to establish a bound in the new effects coming from $e_1$ and $d_1$. In order to avoid cross-interactions from binary systems with multiple spinning sources and given the fact that the current measurements for the masses and gravitational redshifts of isolated neutron stars do not provide independent quantities \cite{Tang:2019ign}, we focus on the Sirius B white dwarf, whose Doppler shift velocity $v=c\,z$ associated with the gravitational redshift has been measured by Balmer line techniques, finding \cite{joyce2018gravitational,Barstow:2005mx}
\begin{eqnarray}
v_{\rm obs, Sirius B}=(80.65\pm 0.77)\, \Big[\frac{\textrm{km}}{s}\Big]\,.\label{vobs}
\end{eqnarray}
Then, with the values for the mass $M= 1.018M_{\odot}$ and radius $R=8.098 \cdot 10^{-3} R_{\odot}$~\cite{bond2017sirius} where $R_{\odot}=695,700\, \textrm{km}$ is the radius of the Sun, the predicted GR contribution of the Doppler shift velocity turns out to be
\begin{eqnarray}\label{vGR}
v_{\rm GR, Sirius B}=80.0464\, \Big[\frac{\textrm{km}}{s}\Big]\,.
\end{eqnarray}

Accordingly, comparing the gravitational Doppler shift velocity~\eqref{vGR} provided by GR and the one related to the gravitational redshift~\eqref{redshift} of our model with the observed value~\eqref{vobs}, we find that the new contributions coming from torsion and nonmetricity are bounded to be
\begin{eqnarray}
4e_1\kappa_d^2- 2.931\cdot 10^{43} \, \left[J\cdot m\right]\leq d_{1}\kappa_s^2\leq 4e_1\kappa_d^2+
1.016\cdot 10^{43} \, \left[J\cdot m\right]\,.
\end{eqnarray}

Let us now consider the case where the effect of torsion dominates over the contribution of nonmetricity. In fact, due to the presence of a magnetic field in white dwarfs~\cite{ferrario2020magnetic}, it is expected that Sirius B can have sufficiently oriented elementary spins in comparison with an effective dilation charge, therefore, $\kappa_{s,\rm Sirius B} \gg \kappa_{d,\rm Sirius B}$. Assuming the same approximation in Sgr A* and considering the universality of the coupling constant $d_{1}$, we can use the bound (\ref{Sagi}) found from the perihelion shift of the S2 star and constrain the spin charges of Sgr A* and Sirius B as follows
\begin{eqnarray}
1.396\cdot 10^{10}\leq \frac{\kappa_{s,\rm  Sgr A*}}{\kappa_{s,\rm Sirius B}} \leq 1.688\cdot 10^{10} \,.
\end{eqnarray}

To the best of our knowledge, this bound provides the first observational comparison between the spin charges of a supermassive black hole and a degenerate star.

\subsection{Shapiro delay and deflection of light}\label{sec:light}
The Shapiro delay is another observable that has shown to be a useful tool to constrain theories of gravity. It measures the time delay of light caused by the geometry of the space-time, which is increased during the propagation of light with respect to a distant observer. This means to find out the time required for light to travel from the distance of the closest encounter with a central mass $M$, that we will denote as $r_0$, to any distance $r>r_0$\,. Since the quantity $r_0$ represents the minimal distance from the source, it fulfills the condition $\dot{r}(r)\rvert_{r=r_0}=0$, which according to Eq. (\ref{potentialEq}) gives us that the effective potential vanishes $V(r_{0})=0$. By solving this equation for massless particles with $\sigma=0$, one gets the following expression for the root $r_0$\,:
\begin{eqnarray}\label{rois}
r_0=\left(\frac{J}{c\, E}\right)\Psi^{1/2}(r_0)\,.
\end{eqnarray}
Now, if we integrate Eq.~\eqref{potentialEq} and replace $r_0$ by using the above equation, one gets that the time $t(r,r_0)$ required for the light to travel from $r_0$ to $r$ is 
\begin{equation}\label{time}
    t(r,r_0)= \frac{1}{c}\int_{r_0}^r \frac{d\bar{r}}{\Psi(\bar{r})}\bigg[\left(1-\frac{r_0^2 \Psi(\bar{r})}{\bar{r}^2 \Psi(r_0)}\right)\bigg]^{-1/2}\,.
\end{equation}
The Shapiro delay is then related to the retardation of light during this time travel as follows. Let us consider a light pulse at a radius $r_e$ propagating through the space-time to a point $r_0$ of closest encounter with the mass $M$, and then propagates to a mirror at radius $r_{\rm m}$. Thus, the light ray is reflected and come back to the emitter. The Shapiro delay of the trip is then defined as
\begin{equation}
  \Delta  t_{\rm Shapiro}(r_{\rm e},r_{\rm m},r_0)=\frac{1}{2}\left(t(r_{\rm e},r_0)+t(r_{\rm m},r_0)-\sqrt{r_{\rm e}^2-r_0^2}-\sqrt{r_{\rm m}^2-r_0^2} \right)\,.
\end{equation}

The photon time travel~\eqref{time} cannot be analytically integrated for our RN type of solution~\eqref{RN}. However, one can consider the corresponding effect at large distances (see Eq.~\eqref{approx}), which essentially means that the gravitational potential is sufficiently small. By following a similar approach as before, and assuming that the emitter is at the same position as the mirror (i.e. $r_e=r_m=r$), we find that the Shapiro delay approximated up to fourth order of the mass $M$ and first order of $\kappa_s^2d_1$ and $\kappa_d^2 e_1$ is 
\begin{eqnarray}
    \Delta t_{\rm Shapiro}(r,r_0)  &\approx&
    \frac{GM\sqrt{r^2-r_0^2}}{c^3(r+r_0)} \Big[1-\frac{G M (4 r+5 r_0)}{2 c^2 r_0 (r+r_0)}+\frac{G^2 M^2 \left(60 r^3+157 r^2 r_0+133 r r_0^2+35 r_0^3\right)}{2 c^4 r r_0^2 (r+r_0)^2}\nonumber\\
    &&-\,\frac{G^3 M^3 \left(3200 r^5+9899 r^4 r_0+9196 r^3 r_0^2+532 r^2 r_0^3-2940 r r_0^4-945 r_0^5\right)}{48 c^6 r^2 r_0^3 (r+r_0)^3}\Big]\nonumber\\
    &&+\,\frac{2 GM}{c^3} \log \Big(\frac{\sqrt{r^2-r_0^2}+r}{r_0}\Big)+\frac{15 \pi  G^2 M^2}{4 c^5 r_0}-\frac{15 \pi  G^3 M^3}{4 c^7 r_0^2}+\frac{1095 \pi  G^4 M^4}{32 c^9 r_0^3}+\frac{3 \pi  G \left(4 e_1\kappa_d^2-d_1\kappa_s^2\right)}{4 c^5 r_0}\nonumber\\
    &&-\arctan\left(\frac{r_0}{\sqrt{r^2-r_0^2}}\right) \left(\frac{15 G^2 M^2}{2 c^5 r_0}-\frac{15 G^3 M^3}{2 c^7 r_0^2}+\frac{1095 G^4 M^4}{16 c^9 r_0^3}-\frac{3 G \left(d_1\kappa_s^2-4 e_1\kappa_d^2\right)}{2 c^5 r_0}\right)\,.\label{eq:shap}
\end{eqnarray}
In order to constrain our model in a realistic way, we would need an astrophysical system  such that $\kappa_d$ and/or $\kappa_s$ are different to zero. The Shapiro delay effect has been used in different scenarios such as in pulsars. For example, using the Shapiro delay, it was possible to determine the masses of the pulsar PSR J1614-2230,
which consists of a binary system composed by a neutron star and a white dwarf companion~\cite{demorest2010two}. This analysis was carried out by comparing the delay of pulse arrivals to the Earth from the light coming from PSR J1614-2230
when it passes behind the white dwarf. However, for our case, we cannot use these data since the authors inferred the masses of the binary system from the time signal and not in the other way around. Hence, on account of the lack of additional observations with an independent estimation of the mass components, one cannot use the Shapiro delay measurements as a natural tool to constrain our model.

The deflection of light can be evaluated as the angle which characterises the difference in the trajectory of a light ray after passing around a gravitational system. This ray of light is then measured in the Earth which is located at a very long distance to the source, so that, one needs to integrate the Eq.~\eqref{eq:rphi} from an impact parameter $r_0$ to $r\rightarrow \infty$.  Thus, this effect can be also caused by the presence of a gravitating central mass $M$ with additional charges. If we again label $r_0$ as the position obtained in~\eqref{rois} such that the effective potential vanishes $V(r_0)=0$, and integrate out the azimuthal angle in Eq.~\eqref{eq:rphi} from $r_0$ to $\infty$, the deflection of light is written as~\cite{Bozza:2009yw}
\begin{equation}
     \Delta\varphi    =\pm 2 \int_{r_0}^\infty \frac{r_0\,d\bar r}{\bar{r}^2\Psi(r_0)^{1/2}}\bigg[1-\left(\frac{r_0}{\bar{r}}\right)^2\frac{\Psi(\bar{r})}{\Psi(r_0)}\bigg]^{-1/2}-\pi\,.\label{phiis}
\end{equation}
The above expression can be analysed by numerical computations (see \cite{Zamani:2020tsm,Zamani:2021hyx} as an example concerning the particular contribution of torsion to this effect). Under the approximation shown in~\eqref{approx}, we find that the deflection of light for the solution~\eqref{RN} acquires the analytical expression
\begin{equation}
     \Delta \varphi\approx  \frac{4 G M}{c^2 r_0}+ \frac{ G^2 M^2}{4 c^4 r_0^2}(15 \pi -16)+\frac{3 \pi  G}{4 c^4 r_0^2} \left(4 e_1 \kappa_d^2-d_1 \kappa_s^2\right)\,.
\end{equation}
The first two terms describe the deflection of light provided by the mass as in the standard case of the Schwarzschild solution, whereas the last one constitutes the correction provided by the hypermomentum charges to this effect.

\section{Conclusions}\label{sec:conclusions}

In the geometric scheme of MAG, not only an energy-momentum tensor of matter arises as source of curvature, but also a hypermomentum density tensor which operates as source of torsion and nonmetricity. For the special case of Weyl-Cartan geometry it splits into spin and dilation currents, which carry their own charges and provide a RN type of solution \eqref{RN} to the field equations of MAG. Following these lines, we compute the dynamical corrections provided by both spin and dilation charges in the main classical tests of gravity, namely the photon sphere, perihelion shift, gravitational redshift, Shapiro delay and deflection of light. Since the aforementioned exact solution can be assumed to describe the gravitational field around a black hole or a compact star formed in a metric-affine regime with scale invariance, we are able to constrain such corrections with recent observations of the S2 star around the supermassive black hole Sgr A* and Sirius B. Indeed, the well-established premise that primordial black holes formed from nonbaryonic matter at very high energies in the early universe may constitute the seeds of supermassive black holes located at the center of massive galaxies places these objects as leading candidates to test the gravitational effects of MAG \cite{Bernal:2017nec}. In addition, the core remnant of white dwarfs such as Sirius B supported by electron degeneracy may exhibit a spin alignment induced by sufficiently intense dipolar magnetic fields, whose range can lie in any case below the current level of experimental accuracy \cite{ferrario2020magnetic}, and accordingly constitute a viable macroscopic source of torsion. In this regard, whether or not degenerate stars can also carry a dilation charge still remains as an open-ended question.

Thereby, we evaluate the contributions of the mass and the hypermomentum charges of the solution to the relativistic perihelion precession of S2 around Sgr A* and to the gravitational redshift in Sirius B up to a fourth order expansion of these parameters. Interestingly, we constrain such contributions with the observed quantities and find a branch compatible with a negative effective correction of the RN geometry. This possibility involves the absence of an inner Cauchy horizon in the solution, which in the Einstein-Maxwell model of GR has shown to be a pathological hypersurface, due to the loss of predictability from initial conditions, and the existence of mass-inflation and kink instabilities \cite{Poisson:1990eh,Maeda:2005yd}. Indeed, even though the field strength tensors of the torsion and nonmetricity fields show a similar Coulomb-like pattern as the electromagnetic fields of the Einstein-Maxwell model, they are sourced by different charges. In particular, the existing correspondence between torsion and the intrinsic angular momentum of matter allows their dynamical effects to be propagated by irreducible modes with different parity, as it is the case in the present model \cite{Bahamonde:2020fnq}. Additional analyses considering this branch as well as the structure and the gravitational stability of the solution are currently underway.

It is worthwhile to stress a general issue to isolate the gravitational redshift from the Doppler shifted due to random stellar motions along the line of observation \cite{chandra2020gravitational}. In the case of Sirius B, this degeneracy can be overcome constraining the radial component of the apparent velocity, but for Sgr A* it has not been solved yet due to the fact that the effect of the relative motion between the observer and Sgr A* is too small to decompose the total redshift~\cite{Abuter:2018drb}. For this reason, one cannot use the mentioned measurement of the redshift in Sgr A* to constrain our model.

On the other hand, we also derive the corresponding corrections to the photon sphere, Shapiro delay and deflection of light provided by torsion and nonmetricity, but again one cannot use yet any observation of astrophysical sources as the ones considered in our work to obtain realistic constraints for these effects. For example for the Shapiro delay and the deflection of light, the astronomy of pulsars is mostly devoted on measurements from the companion astrophysical object rather than the observables related to the neutron star (see~\cite{demorest2010two} as an example). The photon sphere casts the shadow of the black hole and the only known observations for this effect is the one obtained by The Event Horizon Telescope collaboration for M87*~\cite{Akiyama:2019cqa}. We showed that only the size of the shadow provided by our solution is modified, but its shape remains the same. It would be interesting to use these effects in the future whenever these observations are carried out. 

Further constraints to achieve a complete characterisation of the torsion and nonmetricity tensors in MAG require supplementary analyses and observations of the effects of these quantities on test bodies with intrinsic hypermomentum \cite{Puetzfeld:2007hr,Hehl:2013qga}. In this sense, current observations of binary pulsars with multiple spinning sources, such as the double pulsar PSR J0737-3039 or the milisecond pulsar PSR J0437-4715 with a white dwarf companion, must also be considered taking into account the additional Lorentz-like forces yielded by the torsion and nonmetricity tensors, according to the general equations of motion obtained in (\ref{eqsmotion}). In any case, the fact that most of the neutron star-white dwarf binary systems contain very circular orbits as a result of tidal dissipation turns out to be another experimental issue for the measurement of certain relativistic orbital parameters, such as the perihelion shift computed in this work \cite{Wex:2014nva,phinney1992pulsars}. In the same way, the corresponding corrections for ordinary stars or planets around neutron stars are so small to be undetectable.

At the Earth scale, the microscopic signatures of torsion on the energy levels of minimally and nonminimally coupled Dirac spinors have been tested \cite{audretsch1983neutron,Lammerzahl:1997wk,Obukhov:2014fta,Kostelecky:2007kx,Heckel:2008hw,Lehnert:2013jsa}, whereas future tests based on the precession of ferromagnetic gyroscopes have also been successfully proposed \cite{Fadeev:2020gjk}. In this sense, on account of the purely quantum nature of the intrinsic hypermomentum of matter, the performance of new experiments sufficiently sensitive to measure the splitting of the fermionic energy levels in strong gravitational regimes opens up promising prospects for testing post-Riemannian geometry \cite{Cabral:2021dfe}. In particular, the existence of a dynamical axial mode of torsion allows the evaluation of such signatures in Dirac fermions by means of our solution without the requirement of extra nonminimal couplings in the torsion sector. Further research following these lines will be addressed in future works.

\bigskip
\bigskip
\noindent
\section*{Acknowledgements}
The authors would like to thank Clifford M. Will for useful discussions. S.B. was supported by the Estonian Research Council grants PRG356 ``Gauge Gravity"  and by the European Regional Development Fund through the Center of Excellence TK133 ``The Dark Side of the Universe". J.G.V. was supported by the European Regional Development Fund and the programme Mobilitas Pluss (Grant No. MOBJD541). 
\newpage

\appendix
\section{Field equations and conservation law of the canonical energy-momentum tensor}\label{sec:AppFieldEqs}

The variation of the Lagrangian (\ref{LagrangianIrreducible}) provides the following tensor quantities depending on the curvature, torsion and nonmetricity tensors:
\begin{eqnarray}
X1_{\mu}\,^{\nu} &=& \frac{2c^4}{G}\,G_{\mu}\,^{\nu}+16\pi\tilde{\mathcal{L}}\,\delta_{\mu}\,^{\nu}+3\left(1-2a_{2}\right)\left(g_{\mu\rho}\nabla_{\lambda}T^{\left[\lambda\nu\rho\right]}+K_{\lambda\rho\mu}T^{\left[\lambda\nu\rho\right]}\right)-4e_{1}\tilde{R}^{\lambda}\,_{\lambda\sigma\mu}\tilde{R}^{\rho}\,_{\rho}\,^{\sigma\nu}
\nonumber\\
&&+2d_{1}\left(\tilde{R}^{\nu}\,_{\lambda\rho\mu}\tilde{R}^{\left[\lambda\rho\right]}+\tilde{R}_{\lambda}\,^{\nu}\,_{\mu\rho}\hat{R}^{\left[\lambda\rho\right]}+\tilde{R}_{\lambda\mu}\tilde{R}^{\left[\nu\lambda\right]}+\hat{R}_{\lambda\mu}\hat{R}^{\left[\nu\lambda\right]}\right)
\nonumber\\
&&+\frac{1}{2}\tilde{R}_{\lambda\rho\sigma\mu}
\Bigl[
d_{1}\left(4\tilde{R}^{\left[\nu\sigma\right]\lambda\rho}-2\tilde{R}^{\left[\lambda\rho\right]\nu\sigma}-\tilde{R}^{\left[\lambda\nu\right]\rho\sigma}-\tilde{R}^{\left[\rho\sigma\right]\lambda\nu}-\tilde{R}^{\left[\rho\nu\right]\sigma\lambda}-\tilde{R}^{\left[\sigma\lambda\right]\rho\nu}\right)
\Bigr.\,
\nonumber\\
\Bigl.
&&+8e_{2}\left(\tilde{R}^{\left(\rho\nu\right)\lambda\sigma}-\tilde{R}^{\left(\lambda\sigma\right)\rho\nu}+\tilde{R}^{\left(\lambda\nu\right)\rho\sigma}-\tilde{R}^{\left(\rho\sigma\right)\lambda\nu}\right)
\Bigr]\,,
\end{eqnarray}
\begin{eqnarray}
X2^{\lambda\mu\nu}&=&2d_{1}
\Bigl[
\nabla_{\rho}\left(g^{\mu\nu}\tilde{R}^{\left[\lambda\rho\right]}-g^{\lambda\nu}\hat{R}^{\left[\mu\rho\right]}+g^{\lambda\rho}\hat{R}^{\left[\mu\nu\right]}-g^{\mu\rho}\tilde{R}^{\left[\lambda\nu\right]}\right)+\Bigl(K^{\rho\mu}\,_{\rho}+L^{\rho\mu}\,_{\rho}\Bigr)\tilde{R}^{\left[\lambda\nu\right]}-\Bigl(K^{\rho\lambda}\,_{\rho}+L^{\rho\lambda}\,_{\rho}\Bigr)\hat{R}^{\left[\mu\nu\right]}
\Bigr.
\nonumber\\
\Bigl.
&&+\Bigl(K^{\nu\lambda}\,_{\rho}+L^{\nu\lambda}\,_{\rho}\Bigr)\hat{R}^{\left[\mu\rho\right]}-\Bigl(K^{\nu\mu}\,_{\rho}+L^{\nu\mu}\,_{\rho}\Bigr)\tilde{R}^{\left[\lambda\rho\right]}+\Bigl(K^{\mu}\,_{\rho}\,^{\lambda}+L^{\mu}\,_{\rho}\,^{\lambda}\Bigr)\hat{R}^{\left[\rho\nu\right]}-\Bigl(K^{\lambda}\,_{\rho}\,^{\mu}+L^{\lambda}\,_{\rho}\,^{\mu}\Bigr)\tilde{R}^{\left[\rho\nu\right]}
\Bigr]
\nonumber\\
&&+\frac{1}{2}\Bigl(\nabla_{\rho}+W_{\rho}\Bigr)
\Bigl[
d_{1}\left(4\tilde{R}^{\left[\rho\nu\right]\lambda\mu}-2\tilde{R}^{\left[\lambda\mu\right]\rho\nu}-\tilde{R}^{\left[\mu\nu\right]\lambda\rho}+\tilde{R}^{\left[\lambda\nu\right]\mu\rho}-\tilde{R}^{\left[\lambda\rho\right]\mu\nu}+\tilde{R}^{\left[\mu\rho\right]\lambda\nu}\right)
\Bigr.
\nonumber\\
\Bigl.
&&+8e_{2}\left(\tilde{R}^{\left(\mu\nu\right)\lambda\rho}-\tilde{R}^{\left(\lambda\rho\right)\mu\nu}+\tilde{R}^{\left(\nu\lambda\right)\mu\rho}-\tilde{R}^{\left(\rho\mu\right)\lambda\nu}\right)
\Bigr]
-4e_{1}g^{\lambda \mu}\nabla_{\rho}\tilde{R}_{\sigma}\,^{\sigma\rho\nu}-3\left(1-2a_{2}\right)T^{\left[\lambda\mu\nu\right]}
\nonumber\\
&&+\frac{1}{2}\Bigl(K^{\lambda}\,_{\sigma\rho}+L^{\lambda}\,_{\sigma\rho}\Bigr)
\Bigl[
d_{1}\left(4\tilde{R}^{\left[\rho\nu\right]\sigma\mu}-2\tilde{R}^{\left[\sigma\mu\right]\rho\nu}-\tilde{R}^{\left[\mu\nu\right]\sigma\rho}+\tilde{R}^{\left[\sigma\nu\right]\mu\rho}-\tilde{R}^{\left[\sigma\rho\right]\mu\nu}+\tilde{R}^{\left[\mu\rho\right]\sigma\nu}\right)
\Bigr.
\nonumber\\
\Bigl.
&&+8e_{2}\left(\tilde{R}^{\left(\mu\nu\right)\sigma\rho}-\tilde{R}^{\left(\sigma\rho\right)\mu\nu}+\tilde{R}^{\left(\nu\sigma\right)\mu\rho}-\tilde{R}^{\left(\rho\mu\right)\sigma\nu}\right)+4d_{1}g^{\mu\nu}\tilde{R}^{\left[\sigma\rho\right]}
\Bigr]
\nonumber\\
&&+\frac{1}{2}\Bigl(K^{\mu}\,_{\sigma\rho}+L^{\mu}\,_{\sigma\rho}\Bigr)
\Bigl[
d_{1}\left(4\tilde{R}^{\left[\rho\nu\right]\lambda\sigma}-2\tilde{R}^{\left[\lambda\sigma\right]\rho\nu}-\tilde{R}^{\left[\sigma\nu\right]\lambda\rho}+\tilde{R}^{\left[\lambda\nu\right]\sigma\rho}-\tilde{R}^{\left[\lambda\rho\right]\sigma\nu}+\tilde{R}^{\left[\sigma\rho\right]\lambda\nu}\right)
\Bigr.
\nonumber\\
\Bigl.
&&+8e_{2}\left(\tilde{R}^{\left(\sigma\nu\right)\lambda\rho}-\tilde{R}^{\left(\lambda\rho\right)\sigma\nu}+\tilde{R}^{\left(\nu\lambda\right)\sigma\rho}-\tilde{R}^{\left(\sigma\rho\right)\lambda\nu}\right)-4d_{1}g^{\lambda\nu}\hat{R}^{\left[\sigma\rho\right]}
\Bigr]\,,
\end{eqnarray}
in such a way that the correspondence with the canonical energy-momentum and hypermomentum tensors gives rise to the field equations:
\begin{eqnarray}\label{appfield_eq1}
X1_{\mu}\,^{\nu} &=& 16\pi\theta_{\mu}\,^{\nu}\,,\\
X2^{\lambda\mu\nu} &=& 16\pi\bigtriangleup^{\lambda\mu\nu},\label{appfield_eq2}
\end{eqnarray}
with $\tilde{\mathcal{L}}$ representing the Lagrangian density composed by the quadratic order corrections.

The computation of the Riemannian divergence of the Expression (\ref{appfield_eq1}) allows the conservation law of the canonical energy-momentum tensor to be obtained as follows:
\begin{eqnarray}
16\pi\nabla_{\nu}\theta_{\mu}\,^{\nu} &=& 3\left(1-2a_{2}\right)\left\{\Big[K_{\lambda\rho\mu}+L_{[\lambda\rho]\mu}\Bigr]\left[\nabla_{\nu}T^{[\lambda\nu\rho]}+\left(K_{\sigma\nu}\,^{\lambda}+L_{[\sigma\nu]}\,^{\lambda}\right)T^{[\sigma\rho\nu]}\right]-\tilde{R}_{\lambda\nu\rho\mu}T^{[\lambda\nu\rho]}\right\}
\nonumber\\
&&+\,\frac{1}{2}\left(4e_{1}+e_{2}\right)\left(\tilde{R}^{\sigma}\,_{\sigma}\,^{\rho\nu}\nabla_{\mu}\tilde{R}^{\lambda}\,_{\lambda\rho\nu}-2\tilde{R}^{\sigma}\,_{\sigma}\,^{\rho\nu}\nabla_{\nu}\tilde{R}^{\lambda}\,_{\lambda\rho\mu}-2\tilde{R}^{\sigma}\,_{\sigma\rho\mu}\nabla_{\nu}\tilde{R}^{\lambda}\,_{\lambda}\,^{\rho\nu}\right)
\nonumber\\
&&+\,d_{1}\Big\{
\nabla_{\nu}\tilde{R}_{\lambda\rho\sigma\mu}
\Bigl[\left(3\tilde{R}^{\left[\nu\sigma\right]\lambda\rho}-\tilde{R}^{\left[\lambda\rho\right]\nu\sigma}\right)-\frac{1}{2}\left(\tilde{R}^{\left[\lambda\nu\right]\rho\sigma}+\tilde{R}^{\left[\rho\sigma\right]\lambda\nu}+\tilde{R}^{\left[\rho\nu\right]\sigma\lambda}+\tilde{R}^{\left[\sigma\lambda\right]\rho\nu}\right)\Bigr]
\Bigr.
\nonumber\\
\Bigl.
&&+\tilde{R}_{\lambda\rho\sigma\mu}
\Bigl[\nabla_{\nu}\left(3\tilde{R}^{\left[\nu\sigma\right]\lambda\rho}-\tilde{R}^{\left[\lambda\rho\right]\nu\sigma}\right)-\frac{1}{2}\nabla_{\nu}\left(\tilde{R}^{\left[\lambda\nu\right]\rho\sigma}+\tilde{R}^{\left[\rho\sigma\right]\lambda\nu}+\tilde{R}^{\left[\rho\nu\right]\sigma\lambda}+\tilde{R}^{\left[\sigma\lambda\right]\rho\nu}\right)\Bigr]
\Bigr.
\nonumber\\
\Bigl.
&&+2\left(\nabla_{\nu}\tilde{R}^{\nu}\,_{\lambda\rho\mu}\tilde{R}^{\left[\lambda\rho\right]}+\nabla_{\nu}\tilde{R}_{\lambda\mu}\tilde{R}^{\left[\nu\lambda\right]}+\tilde{R}^{\nu}\,_{\lambda\rho\mu}\nabla_{\nu}\tilde{R}^{\left[\lambda\rho\right]}+\tilde{R}_{\lambda\mu}\nabla_{\nu}\tilde{R}^{\left[\nu\lambda\right]}+\tilde{R}^{\left[\sigma\nu\right]}\nabla_{\mu}\tilde{R}_{\left[\sigma\nu\right]}\right.
\Bigr.
\nonumber\\
\Bigl.
&&\left.+\nabla_{\nu}\tilde{R}_{\lambda}\,^{\nu}\,_{\mu\rho}\hat{R}^{\left[\lambda\rho\right]}+\nabla_{\nu}\hat{R}_{\lambda\mu}\hat{R}^{\left[\nu\lambda\right]}+\tilde{R}_{\lambda}\,^{\nu}\,_{\mu\rho}\nabla_{\nu}\hat{R}^{\left[\lambda\rho\right]}+\hat{R}_{\lambda\mu}\nabla_{\nu}\hat{R}^{\left[\nu\lambda\right]}+\hat{R}^{\left[\sigma\nu\right]}\nabla_{\mu}\hat{R}_{\left[\sigma\nu\right]}\right)
\Bigr.
\nonumber\\
\Bigl.
&&+\frac{1}{2}\,\nabla_{\mu}\tilde{R}_{\lambda\rho\sigma\nu}
\left(3\tilde{R}^{\left[\sigma\nu\right]\lambda\rho}-\tilde{R}^{\left[\lambda\rho\right]\sigma\nu}-\tilde{R}^{\left[\lambda\sigma\right]\rho\nu}-\tilde{R}^{\left[\sigma\rho\right]\lambda\nu}\right)\Bigr\}\,.
\end{eqnarray}

By taking into account the Expression (\ref{appfield_eq2}) and the identity $\nabla_{[\rho|}\tilde{R}^{\lambda}\,_{\lambda|\mu\nu]}=0$ of the homothetic curvature tensor:
\begin{eqnarray}
16\pi\nabla_{\nu}\theta_{\mu}\,^{\nu} &=& 3\left(1-2a_{2}\right)\Big[K_{\lambda\rho\mu}+L_{[\lambda\rho]\mu}\Bigr]\left[\nabla_{\nu}T^{[\lambda\nu\rho]}+\left(K_{\sigma\nu}\,^{\lambda}+L_{[\sigma\nu]}\,^{\lambda}\right)T^{[\sigma\rho\nu]}\right]-16\pi\tilde{R}_{\lambda\rho\sigma\mu}\bigtriangleup^{\rho\lambda\sigma}
\nonumber\\
&&+\,d_{1}\Big\{
\nabla_{\nu}\tilde{R}_{\lambda\rho\sigma\mu}
\Bigl[\left(3\tilde{R}^{\left[\nu\sigma\right]\lambda\rho}-\tilde{R}^{\left[\lambda\rho\right]\nu\sigma}\right)-\frac{1}{2}\left(\tilde{R}^{\left[\lambda\nu\right]\rho\sigma}+\tilde{R}^{\left[\rho\sigma\right]\lambda\nu}+\tilde{R}^{\left[\rho\nu\right]\sigma\lambda}+\tilde{R}^{\left[\sigma\lambda\right]\rho\nu}\right)\Bigr]
\Bigr.
\nonumber\\
\Bigl.
&&-\tilde{R}_{\lambda\rho\sigma\mu}
\Bigl[W_{\nu}\left(3\tilde{R}^{\left[\nu\sigma\right]\lambda\rho}-\tilde{R}^{\left[\lambda\rho\right]\nu\sigma}\right)-\frac{1}{2}W_{\nu}\left(\tilde{R}^{\left[\lambda\nu\right]\rho\sigma}+\tilde{R}^{\left[\rho\sigma\right]\lambda\nu}+\tilde{R}^{\left[\rho\nu\right]\sigma\lambda}+\tilde{R}^{\left[\sigma\lambda\right]\rho\nu}\right)\Bigr]
\Bigr.
\nonumber\\
\Bigl.
&&-N^{\lambda}\,_{\omega\nu}\tilde{R}_{\lambda\rho\sigma\mu}
\Bigl[\left(3\tilde{R}^{\left[\nu\sigma\right]\omega\rho}-\tilde{R}^{\left[\omega\rho\right]\nu\sigma}\right)-\frac{1}{2}\left(\tilde{R}^{\left[\omega\nu\right]\rho\sigma}+\tilde{R}^{\left[\rho\sigma\right]\omega\nu}+\tilde{R}^{\left[\rho\nu\right]\sigma\omega}+\tilde{R}^{\left[\sigma\omega\right]\rho\nu}\right)+2g^{\sigma\rho}\hat{R}^{\left[\omega\nu\right]}\Bigr]
\Bigr.
\nonumber\\
\Bigl.
&&-N^{\rho}\,_{\omega\nu}\tilde{R}_{\lambda\rho\sigma\mu}
\Bigl[\left(3\tilde{R}^{\left[\nu\sigma\right]\lambda\omega}-\tilde{R}^{\left[\lambda\omega\right]\nu\sigma}\right)-\frac{1}{2}\left(\tilde{R}^{\left[\lambda\nu\right]\omega\sigma}+\tilde{R}^{\left[\omega\sigma\right]\lambda\nu}+\tilde{R}^{\left[\omega\nu\right]\sigma\lambda}+\tilde{R}^{\left[\sigma\lambda\right]\omega\nu}\right)-2g^{\lambda\sigma}\tilde{R}^{\left[\omega\nu\right]}\Bigr]
\Bigr.
\nonumber\\
\Bigl.
&&+2\left(\nabla_{\nu}\tilde{R}^{\nu}\,_{\lambda\rho\mu}\tilde{R}^{\left[\lambda\rho\right]}+\nabla_{\nu}\tilde{R}_{\lambda\mu}\tilde{R}^{\left[\nu\lambda\right]}+\tilde{R}^{\left[\sigma\nu\right]}\nabla_{\mu}\tilde{R}_{\left[\sigma\nu\right]}+\nabla_{\nu}\tilde{R}_{\lambda}\,^{\nu}\,_{\mu\rho}\hat{R}^{\left[\lambda\rho\right]}+\nabla_{\nu}\hat{R}_{\lambda\mu}\hat{R}^{\left[\nu\lambda\right]}+\hat{R}^{\left[\sigma\nu\right]}\nabla_{\mu}\hat{R}_{\left[\sigma\nu\right]}\right)
\Bigr.
\nonumber\\
\Bigl.
&&+\tilde{R}_{\lambda\rho\sigma\mu}
\left(N^{\nu\lambda}\,_{\nu}\tilde{R}^{\left[\rho\sigma\right]}-N^{\nu\rho}\,_{\nu}\hat{R}^{\left[\lambda\sigma\right]}+N^{\sigma\rho}\,_{\nu}\hat{R}^{\left[\lambda\nu\right]}-N^{\sigma\lambda}\,_{\nu}\tilde{R}^{\left[\rho\nu\right]}+N^{\lambda}\,_{\nu}\,^{\rho}\hat{R}^{\left[\nu\sigma\right]}-N^{\rho}\,_{\nu}\,^{\lambda}\tilde{R}^{\left[\nu\sigma\right]}\right)
\Bigr.
\nonumber\\
\Bigl.
&&+\frac{1}{2}\,\nabla_{\mu}\tilde{R}_{\lambda\rho\sigma\nu}
\left(3\tilde{R}^{\left[\sigma\nu\right]\lambda\rho}-\tilde{R}^{\left[\lambda\rho\right]\sigma\nu}-\tilde{R}^{\left[\lambda\sigma\right]\rho\nu}-\tilde{R}^{\left[\sigma\rho\right]\lambda\nu}\right)\Bigr\}\,.
\end{eqnarray}

According to the second Bianchi identity in a Weyl-Cartan space-time, the curvature, torsion and nonmetricity tensors satisfy:
\begin{equation}
\tilde{\nabla}_{[\mu|}\tilde{R}_{\lambda\rho|\sigma\nu]} =  T^{\omega}\,_{[\mu\sigma|}\tilde{R}_{\lambda\rho\omega|\nu]}+W_{[\mu|}\tilde{R}_{\lambda\rho|\sigma\nu]}\,,
\end{equation}
where $\tilde{\nabla}_{\mu}\tilde{R}_{\lambda\rho\sigma\nu}=\nabla_{\mu}\tilde{R}_{\lambda\rho\sigma\nu}-N^{\omega}\,_{\lambda\mu}\tilde{R}_{\omega\rho\sigma\nu}-N^{\omega}\,_{\rho\mu}\tilde{R}_{\lambda\omega\sigma\nu}-N^{\omega}\,_{\sigma\mu}\tilde{R}_{\lambda\rho\omega\nu}-N^{\omega}\,_{\nu\mu}\tilde{R}_{\lambda\rho\sigma\omega}$. Then, the second order corrections depending on the curvature tensor present in the divergence of the canonical energy-momentum tensor can be written in terms of the covariant derivative with respect to the general affine connection and rearranged by the second Bianchi identity, which simplifies the expression above in the following way:
\begin{eqnarray}
16\pi\nabla_{\nu}\theta_{\mu}\,^{\nu} &=& 3\left(1-2a_{2}\right)\Big[K_{\lambda\rho\mu}+L_{[\lambda\rho]\mu}\Bigr]\left[\nabla_{\nu}T^{[\lambda\nu\rho]}+\left(K_{\sigma\nu}\,^{\lambda}+L_{[\sigma\nu]}\,^{\lambda}\right)T^{[\sigma\rho\nu]}\right]-16\pi\tilde{R}_{\lambda\rho\sigma\mu}\bigtriangleup^{\rho\lambda\sigma}
\nonumber\\
&&+\,2d_{1}\Bigl[
\tilde{R}^{\left[\lambda\rho\right]}\left(\nabla_{\nu}\tilde{R}^{\nu}\,_{\lambda\rho\mu}+\nabla_{\lambda}\tilde{R}_{\rho\mu}+\nabla_{\mu}\tilde{R}_{\left[\lambda\rho\right]}+N^{\nu}\,_{\lambda\rho}\tilde{R}_{\nu\mu}-N^{\nu}\,_{\lambda\sigma}\tilde{R}^{\sigma}\,_{\nu\rho\mu}-N^{\sigma}\,_{\nu\rho}\tilde{R}^{\nu}\,_{\lambda\sigma\mu}-N^{\nu}\,_{\sigma\nu}\tilde{R}^{\sigma}\,_{\rho\lambda\mu}\right)
\Bigr.
\nonumber\\
\Bigl.
&&+\hat{R}^{\left[\lambda\rho\right]}\left(\nabla_{\nu}\tilde{R}_{\lambda}\,^{\nu}\,_{\mu\rho}+\nabla_{\lambda}\hat{R}_{\rho\mu}+\nabla_{\mu}\hat{R}_{\left[\lambda\rho\right]}+N^{\nu}\,_{\lambda\rho}\hat{R}_{\nu\mu}-N^{\nu}\,_{\lambda\sigma}\tilde{R}_{\nu}\,^{\sigma}\,_{\mu\rho}-N^{\sigma}\,_{\nu\rho}\tilde{R}_{\lambda}\,^{\nu}\,_{\mu\sigma}-N^{\nu}\,_{\sigma\nu}\tilde{R}_{\rho}\,^{\sigma}\,_{\mu\lambda}\right)
\Bigr.
\nonumber\\
\Bigl.
&&+\frac{1}{2}\left(K_{\lambda\rho\mu}+L_{[\lambda\rho]\mu}\right)\tilde{R}^{\left[\lambda\omega\right]\sigma\nu}\left(\tilde{R}_{\left[\sigma\omega\right]\nu}\,^{\rho}-2\tilde{R}_{\left[\sigma\nu\right]\omega}\,^{\rho}\right)\Bigr]\,.
\end{eqnarray}

The same procedure can be considered for the simplification of the second order corrections depending on the Ricci and co-Ricci tensors, which satisfy their own contracted versions of the second Bianchi identity:
\begin{equation}
\tilde{\nabla}_{\nu}\tilde{R}^{\nu}\,_{\lambda\rho\mu}+\tilde{\nabla}_{\mu}\tilde{R}_{\lambda\rho}-\tilde{\nabla}_{\rho}\tilde{R}_{\lambda\mu} =  T^{\sigma}\,_{\nu\rho}\tilde{R}^{\nu}\,_{\lambda\sigma\mu}+T^{\sigma}\,_{\mu\nu}\tilde{R}^{\nu}\,_{\lambda\sigma\rho}+T^{\sigma}\,_{\mu\rho}\tilde{R}_{\lambda\sigma}\,,
\end{equation}
\begin{equation}
\tilde{\nabla}_{\nu}\tilde{R}_{\lambda}\,^{\nu}\,_{\mu\rho}+\tilde{\nabla}_{\mu}\hat{R}_{\lambda\rho}-\tilde{\nabla}_{\rho}\hat{R}_{\lambda\mu} =  T^{\sigma}\,_{\nu\rho}\tilde{R}_{\lambda}\,^{\nu}\,_{\mu\sigma}+T^{\sigma}\,_{\mu\nu}\tilde{R}_{\lambda}\,^{\nu}\,_{\rho\sigma}+T^{\sigma}\,_{\mu\rho}\hat{R}_{\lambda\sigma}\,,
\end{equation}
and give rise to the following compact expression:
\begin{eqnarray}
16\pi\nabla_{\nu}\theta_{\mu}\,^{\nu} &=& 3\left(1-2a_{2}\right)\Big[K_{\lambda\rho\mu}+L_{[\lambda\rho]\mu}\Bigr]\left[\nabla_{\nu}T^{[\lambda\nu\rho]}+\left(K_{\sigma\nu}\,^{\lambda}+L_{[\sigma\nu]}\,^{\lambda}\right)T^{[\sigma\rho\nu]}\right]-16\pi\tilde{R}_{\lambda\rho\sigma\mu}\bigtriangleup^{\rho\lambda\sigma}
\nonumber\\
&&+\,2d_{1}\left(K_{\lambda\rho\mu}+L_{[\lambda\rho]\mu}\right)\left[\tilde{R}^{\rho}\,_{\sigma\omega}\,^{\lambda}\tilde{R}^{\left[\sigma\omega\right]}+\tilde{R}^{\lambda}\,_{\sigma}\tilde{R}^{\left[\rho\sigma\right]}+\tilde{R}_{\sigma}\,^{\rho\lambda}\,_{\omega}\hat{R}^{\left[\sigma\omega\right]}+\hat{R}^{\lambda}\,_{\sigma}\hat{R}^{\left[\rho\sigma\right]}\right]
\nonumber\\
&&+\,d_{1}\left(K_{\lambda\rho\mu}+L_{[\lambda\rho]\mu}\right)\tilde{R}^{\left[\lambda\omega\right]\sigma\nu}\left(\tilde{R}_{\left[\sigma\omega\right]\nu}\,^{\rho}-2\tilde{R}_{\left[\sigma\nu\right]\omega}\,^{\rho}\right)\,.
\end{eqnarray}
In any case, the existing correspondence between the canonical energy-momentum tensor and the geometric corrections provided by the field equation (\ref{appfield_eq1}) displays an antisymmetric component of the canonical energy-momentum tensor:
\begin{eqnarray}
\theta^{\left[\mu\nu\right]} &=& 3\left(1-2a_{2}\right)\left\{\nabla_{\lambda}T^{[\mu\lambda\nu]}+\frac{1}{2}\left[\left(K_{\lambda\rho}\,^{\mu}+L_{[\lambda\rho]}\,^{\mu}\right)T^{[\lambda\nu\rho]}-\left(K_{\lambda\rho}\,^{\nu}+L_{[\lambda\rho]}\,^{\nu}\right)T^{[\lambda\mu\rho]}\right]\right\}
\nonumber\\
&&+\,\frac{d_{1}}{4}\Bigl[
\tilde{R}_{\lambda\rho\sigma}\,^{\mu}\left(4\tilde{R}^{\left[\nu\sigma\right]\lambda\rho}+\tilde{R}^{\left[\lambda\nu\right]\sigma\rho}+\tilde{R}^{\left[\rho\nu\right]\lambda\rho}\right)-\tilde{R}_{\lambda\rho\sigma}\,^{\nu}\left(4\tilde{R}^{\left[\mu\sigma\right]\lambda\rho}+\tilde{R}^{\left[\lambda\mu\right]\sigma\rho}+\tilde{R}^{\left[\rho\mu\right]\lambda\rho}\right)
\Bigr.
\nonumber\\
\Bigl.
&&+8\left(\tilde{R}^{\left[\lambda\rho\right]}\tilde{R}^{[\nu}\,_{\lambda\rho}\,^{\mu]}+\hat{R}^{\left[\lambda\rho\right]}\tilde{R}_{\lambda}\,^{[\nu\mu]}\,_{\rho}\right)+4\left(\tilde{R}_{\lambda}\,^{[\mu}\tilde{R}^{\nu]\lambda}+\hat{R}_{\lambda}\,^{[\mu}\hat{R}^{\nu]\lambda}\right)
\Bigr]\,,
\end{eqnarray}
which allows us to obtain the final expression of the conservation law of the canonical energy-momentum tensor:
\begin{equation}
\nabla_{\nu}\theta_{\mu}\,^{\nu}+\left(K_{\lambda\rho\mu}+L_{[\lambda\rho]\mu}\right)\theta^{\left[\rho\lambda\right]}+\tilde{R}_{\lambda\rho\sigma\mu}\bigtriangleup^{\rho\lambda\sigma}=0\,.
\end{equation}

\bibliographystyle{utphys}
\bibliography{references}

\end{document}